\documentclass[12pt,english]{article}
\usepackage[T1]{fontenc}
\usepackage[latin1]{inputenc}
\usepackage{geometry}
\usepackage{graphicx}
\usepackage{setspace}
\usepackage{amssymb}

\makeatletter

\providecommand{\tabularnewline}{\\}

\newcommand{\lyxaddress}[1]{
\par {\raggedright #1
\vspace{1.4em}
\noindent\par}
}
\newenvironment{lyxlist}[1]
{\begin{list}{}
{\settowidth{\labelwidth}{#1}
 \setlength{\leftmargin}{\labelwidth}
 \addtolength{\leftmargin}{\labelsep}
 }}
{\end{list}}

\usepackage{a4wide}
\usepackage{babel}
\makeatother
\begin{document}

\title{\textbf{\huge An approximation of the ideal scintillation detector
line shape with a generalized gamma distribution}}

\author{O. Ju. Smirnov }

\maketitle
An approximation of the real line shape of a scintillation detector
with a generalized gamma distribution is proposed. The approximation
describes the ideal scintillation line shape  better than the conventional
normal distribution. Two parameters of the proposed function are uniquely
defined by the first two moments of the detector response.

\lyxaddress{Joint Institute for Nuclear Research, Joliot Curie 6, 141980, Dubna
Moscow region, Russia; osmirnov@jinr.ru}


\section{Introduction}

It is known that the response of a scintillation detector can't be
approximated by a symmetric shape since the line skewness is not zero
\cite{Breitenberger} (see also discussion below). An example of the
situation where the deviations of the line shape from a gaussian can
lead to systematic errors is the search for the effects on the tail
of beta-spectra: smearing of the spectrum due to the detector's finite
resolution provides a stronger underlying background in comparison
to what one would expect in the case of a gaussian line shape. 

The purpose of this work is to provide a simple analytical expression
for the asymmetrical shape approximating the corresponding ideal scintillation
detector response for average scintillation intensity counting from
tens to hundreds of registered photoelectrons.

\section{Ideal scintillation detector }

The statistical properties of a scintillation detector response were
studied by Breitenberg \cite{Breitenberger} and independently by
Wright \cite{Wright}. They showed that the relative variance $v_{Q}\equiv\frac{\sigma_{Q}^{2}}{\mu^{2}}$
of the scintillation detector pulse height is:

\begin{equation}
v_{Q}=v_{T}+(1+v_{T})(v_{n}-\frac{1}{\overline{n}})+\frac{1+v_{1}}{\mu},\label{VarQ}\end{equation}

where $v_{T}$ is the relative variance of the photons transfer efficiency,
$\mu$ is the mean signal registered at the photomultiplier (PMT)
anode, measured in photoelectrons (p.e.), $\overline{n}$ is the mean
number of photons produced in a scintillation event and $v_{n}$ is
a relative variance of the number of photons (which reduces to $\frac{1}{\overline{n}}$
in the case of the normal or Poisson variance), and $v_{1}=\left(\frac{\sigma_{1}}{q_{1}}\right)^{2}$
is a relative variance of the single photoelectron response (s.e.r.)
of the photomultiplier ($q_{1}$ and $\sigma_{1}$ are mean position
and variance of the single p.e. peak). 

We will consider an ideal detector with the following features:

\begin{enumerate}
\item anode signal for a single registered photoelectron is described by
normal distribution;
\item the photoelectrons are registered statistically independent;
\item the number of registered photoelectrons (p.e.) $n$ for a monoenergetic
source with a mean number of registered p.e. $\mu$, follows a Poisson
distribution, $P(n)=\frac{\mu^{n}}{n!}e^{-\mu}$;
\item intrinsic line-width of the scintillator is negligible, the variance
of the number of scintillation photons is normal;
\item the detector is spatially uniform, i.e. events with the same energy
produce identical responses on the average at any position inside
the detector;
\item noises in the system are negligible.
\end{enumerate}
As it will be shown below, condition (1) is essential only when registering
on the average small numbers of p.e. in an event, $\mu\lesssim8$.
Condition (2) is satisfied practically automatically in the case of
detector with many PMTs working in single-electron regime, but could
be questionable for a scintillator crystal coupled to a single PMT.
Assumption (3) is natural, but (4) can need further validation in
a real-world detector (especially in the case of the solid-state scintillators).
Condition (5) is difficult to satisfy for large volume detectors,
but in the case of a spatially non- uniform detector it is enough
to introduce an additional parameter $v_{T}$, defined above, to improve
the fit quality. An example of fitting the $^{14}$C beta- decay spectrum
in a large volume non-uniform detector will be given below (see subsection
\ref{Carbon14subsection}).

In \cite{Resolutions} the case of a real scintillation detector with
many PMTs is considered, and it is shown that in the above assumptions
(\ref{VarQ}) reduces to:

\begin{equation}
v_{Q}=\frac{1+\overline{v_{1}}}{\mu},\label{VarQ0}\end{equation}

where $\overline{v_{1}}$ is a relative variance of the single photoelectron
response averaged over all PMTs of the detector. Thus the scintillation
detector consisting of many identical PMTs, surrounding the scintillator
can be considered as one PMT with an extended photocathode. For this
reason the terms {}``PMT'' and the {}``detector'' will not be
distinguished in the following discussion.

If the PMT response (anode output pulse height $q$) to precisely
$n$ photoelectrons is $f_{n}(q)$, and the number of the registered
photoelectrons is distributed according to distribution $P(n)$, then
the PMT response function can be written as $f(q)=\sum P(n)f_{n}$.
The PMT response function here is the probability density function
(p.d.f.), it is normalized to the unity. At the absence of photoelectrons
at the input of the electron multiplier ($n=0$) the PMT is registering
the noise of the system in accordance with the p.d.f. $f_{0}(q)$.
Using the assumption of statistical independence of the registered
photoelectrons one can write the p.d.f. of registering precisely $n$
photoelectrons as a convolution of $n$ independent single-photoelectron
signals $f_{n}=f_{1}\otimes...\otimes f_{1}$. If $f_{1}$ is described
with a normal distribution, then $f_{n}$ follows a normal distribution
as well, with mean $n\cdot q_{1}$ and variance $\sigma_{n}=\sqrt{n}\sigma_{1}$.

With a proper choice of $f_{1}(q)$ function the p.d.f. of the PMT
response can be constructed at any mean scintillation intensity $\mu$:

\begin{equation}
f(q)=\sum_{n=0}P(n)f_{n}(q)=P(0)f_{0}(q)+\sum_{n=1}P(n)f_{n}(q)\otimes f_{0}(q).\label{PMT_pdf}\end{equation}

The Fourier transform of (\ref{PMT_pdf}) gives the characteristic
function:

\begin{equation}
\chi(s)=P(0)\chi_{0}(s)+\sum_{n=1}P(n)\chi_{1}^{n}(s)\chi_{0}(s),\label{FTransform}\end{equation}

where $\chi_{1}(s)$ and $\chi_{0}(s)$ are characteristic functions
of the single photoelectron response and noise, respectively. 

For the case of the Poisson distribution of the probability to register
precisely $n$ p.e. in a scintillation event of mean intensity $\mu$
p.e., the contributions from $n=1,2...$ p.e. can be summed in and
(\ref{FTransform}) can be rewritten in a more compact way:

\begin{equation}
\chi(s)=e^{-\mu}\chi_{0}(s)+\sum_{n=1}\frac{\mu^{n}}{n!}e^{-\mu}\chi_{1}^{n}(s)\chi_{0}(s)=e^{\mu(\chi_{1}(s)-1)}\chi_{0}(s).\label{FTransform2}\end{equation}

The analogous formula can be obtained for the generating function
by using the elementary facts from the theory of branching processes
\cite{Sevastianov}. In fact, omitting the noise term, equation (\ref{FTransform2})
corresponds to a 2-stage cascade device: the photocathode and electrostatic
focusing system providing on the average $\mu$ Poisson-distributed
photoelectrons at the entrance of the electron multiplier with generating
function $G_{2}(s)=e^{\mu(s-1)}$; and the electron multiplier itself
with a single photoelectron response at anode $f_{1}(q)$ with corresponding
generating function $G_{1}(s)$. The resulting generating function
has the same form as (\ref{FTransform2}): $G(s)=G_{2}(G_{1}(s))=e^{\mu(G_{1}(s)-1)}$,
except of the noise term $\chi_{0}(s)$. 

Omitting the noise term, equation (\ref{PMT_pdf}) gets the form $f(x)=\sum_{n=0}P(n)f_{n}(x)$
with characteristic function $\chi(s)=e^{\mu(\chi_{1}(s)-1)}$, which
defines the so called compound Poisson distribution: the probability
distribution of a \char`\"{}Poisson-distributed number\char`\"{} of
independent identically-distributed random variables \cite{Feller}.
In our case the elementary distribution is the s.e.r., and the number
of the independently registered photoelectrons varies in accordance
with Poisson distribution (assumptions 2 and 3).

The inverse transform of (\ref{FTransform2}) in some special cases
of $\chi_{1}(s)$ can be performed analytically, for example, the
case of an exponential single photoelectron response was considered
by Prescott in \cite{Prescott63}.

An example of realistic function $f_{1}(q)$ is shown in Fig.\ref{figure:ser}.
This is the average response observed for the ETL9351 photomultiplier
used in the Borexino detector \cite{BORgen}, the measured mean relative
variance over a set of 2200 PMTs selected for the detector is $v_{1}=0.34$
\cite{2200PMTs}. If the single photoelectron response of PMT and
noise function are known, then formula (\ref{FTransform2}) can be
used to construct the PMT response for any $\mu$ for which the basic
assumptions are valid. The method based on the use of transform (\ref{FTransform2})
has been successfully applied to fit the experimental spectra obtained
with electrostatically focused hybrid photomultiplier tubes for few
registered photoelectrons ($\mu=2.66$ and $\mu=6.36$ p.e.) in \cite{deFatis},
where formula (\ref{FTransform2}) was called \char`\"{}light spectra
sum rule\char`\"{}. 

\begin{figure}
\begin{centering}\includegraphics[width=0.6\paperwidth,height=0.4\paperwidth]{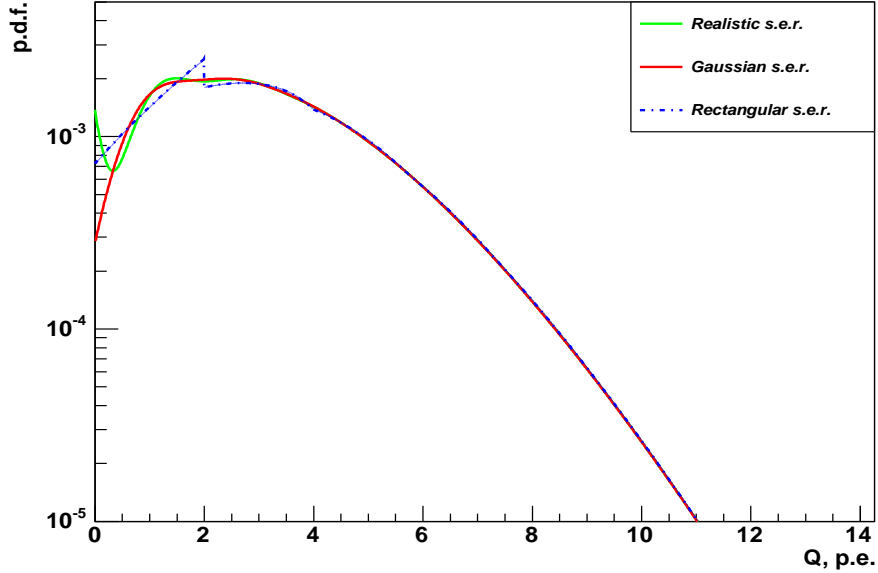}\par\end{centering}

\caption{\label{figure:rectang}Photomultiplier response obtained for 3 different
single electron response functions for the case $\mu=3$ p.e.}
\end{figure}

It should be noted that single photoelectron spectra of the photomultiplier
studied in \cite{deFatis} has a very narrow single p.e. peak, so
that the detector response to $\mu=6.36$ has \char`\"{}fine structure''
peaks around the values corresponding to integer numbers of the registered
charge. In this article we consider a case of $\mu\gg\mu_{0}$ with
$\mu_{0}$ big enough to make the contribution of the first resolved
$n-$fold photoelectron peaks to be negligibly small. The parameter
$\mu_{0}$ can be obtained from the following considerations. The
PMT response to precisely $n$ p.e. (n-fold peak) with increase of
$n$ converges very fast to a normal distribution with $\overline{q}=nq_{1}$
and $\sigma^{2}=n\sigma_{1}^{2}$ as it follows from the central limit
theorem. In practice the PMT response to as low as $n\geq3$ p.e.
can be approximated by a gaussian, see i.e. \cite{Filters}. The (n-1)-fold
and n-fold peaks are not resolved if the half width on the half heights
resolution of the n-th peak is worse than $\frac{1}{2}q_{1}$: $\sqrt{2\mathrm{{ln}}2}\sqrt{n\sigma_{1}^{2}}>\frac{1}{2}q_{1}$,
i.e. $n>\frac{0.18}{v_{1}}$. The contribution of responses from few
photoelectrons decreases very fast with the increase of $\mu$. It
is easy to check that the condition $P(0)+P(1)+P(2)<0.01$ is satisfied
already at $\mu_{0}\simeq8$ p.e. In this case instead of the real
shape $f_{1}(q)$ of the PMT single electron response one can choose
the gaussian approximation for the function $f_{1}(q)$, with mean
$q_{1}$ and variance $\sigma_{1}$ coinciding with the corresponding
parameters of the real-shape function. Indeed, the response functions
for 3 and more p.e. are well approximated by a normal distribution,
and 0,1 and 2 photoelectrons contribute less than 1\% to the total
spectrum (see also Fig.\ref{figure:rectang}).

\begin{center}%
\begin{figure}
\begin{centering}\includegraphics[width=0.6\paperwidth,height=0.4\paperwidth]{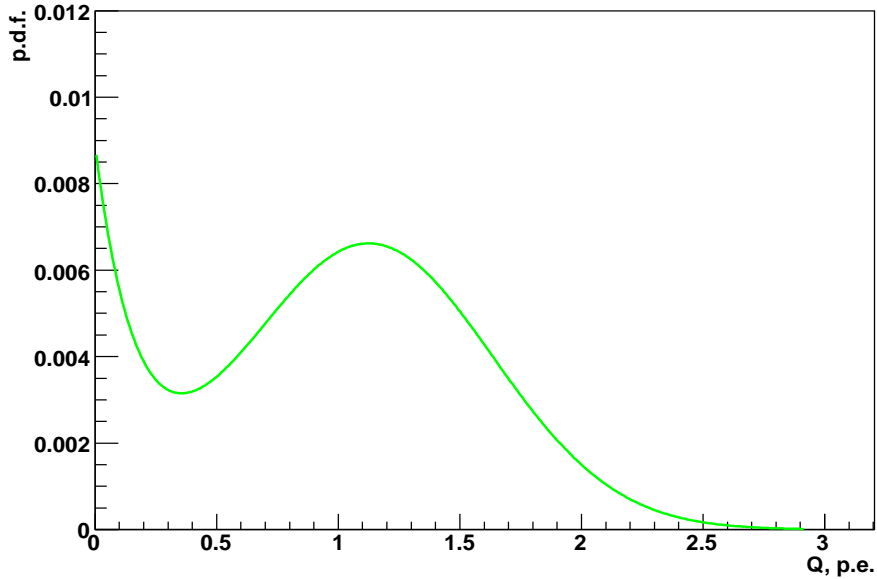}\par\end{centering}

\caption{\label{figure:ser}An example of the single electron response}
\end{figure}
\par\end{center}

In such a way an ideal detector response is described by the inverse
transform of (\ref{FTransform2}) with $\chi_{1}(s)$ corresponding
to the characteristic function of a gaussian with the mean value and
variance of the corresponding single photoelectron response:

\begin{equation}
\chi_{1}(s)=e^{-\frac{1}{2}\sigma_{1}^{2}s^{2}}e^{iq_{1}s}.\label{eq:chi1}\end{equation}

In the following discussion we call the \char`\"{}ideal\char`\"{}
detector response obtained from (\ref{FTransform2}) by using $\chi_{1}(s)$
from (\ref{eq:chi1}), and we let the \char`\"{}real\char`\"{} detector
response to refer to (\ref{FTransform2}) with $\chi_{1}(s)$ obtained
by transforming the real shape of the single photoelectron response.
The difference between the \char`\"{}real\char`\"{} and \char`\"{}ideal\char`\"{}
scintillation response vanishes very fast with the increase of $\mu$
(at $\mu\gtrsim8$ p.e.). We have chosen the gaussian shape for s.e.r.
for convenience, but any appropriate s.e.r. line shape can be used
(with a relative variance that of real s.e.r.). This is illustrated
in Fig.\ref{figure:rectang}, where the theoretical photomultiplier
responses for $\mu=3$ p.e. obtained for 3 different s.e.r. function
(realistic from Fig.\ref{figure:ser}, gaussian and rectangular) with
the same mean value and variance, are plotted. One can see that the
difference is noticeable only at the registered charge $Q<3$ p.e.,
the tail of the PMT response is modeled equally good with the gaussian
and rectangular s.e.r. functions%
\footnote{So, attempts to evaluate the single electron response spectrum at
$\mu\gtrsim1$ seems to be senseless for the PMT spectra with unresolved
s.e.r. ($v_{1}>0.18$), in the best case one can succeed to extract
$q_{1}$ and $v_{1}$values, but not the details of the s.e.r. shape.%
}.

\section{The normal distribution as a limit case for ideal scintillation detector
response}

The ideal detector response converges quickly to the normal distribution
as $\mu$ grows. In fact, the Poisson distribution of the primary
photoelectrons at the input of the electron multiplier converges to
a normal distribution for big $\mu$. The variance in the multiplication
of the photoelectrons arriving at the electron multiplier, for high
$\mu$ values can be considered roughly the same for all possible
values of the registered number of photoelectrons ($\sigma(\mu+\Delta\mu)=\sqrt{\mu}(1+v_{1})+\frac{1}{2}\frac{1+v_{1}}{\sqrt{\mu}}\Delta\mu+..\simeq\sigma(\mu)$).
So, in the big $\mu$ limit the ideal response converges to the convolution
of two gaussian processes which give a normal distribution with the
mean value and variance, respectively: 

\[
\overline{q}=\mu\cdot q_{1},\]

\begin{equation}
\sigma^{2}=(1+v_{1})\cdot\frac{\overline{q}^{2}}{\mu}=(q_{1}^{2}+\sigma_{1}^{2})\mu,\label{eq:MeanAndVar}\end{equation}
coinciding with the values found above considering statistical properties
of the scintillation registration process. We assume that the scale
is calibrated in photoelectrons, i.e. $q_{1}=1$ (otherwise it is
necessary to pass to variable $\frac{q}{q_{1}}$). The characteristic
function for a gaussian p.d.f. is:

\begin{equation}
\chi(s)=e^{-\frac{1}{2}\sigma_{q}^{2}s^{2}}e^{i\overline{q}s}\label{eq:ChiGauss}\end{equation}
and it is apparently different from an ideal shape characteristic
function (\ref{FTransform2}) with $\chi_{1}(s)$ from (\ref{eq:chi1}).
Moreover, one can calculate the moments of the ideal scintillator
response from its generating function:

\begin{equation}
M_{n}=(-i)^{n}\frac{d^{n}\chi(s)}{ds^{n}}\mid_{s=0},\label{MomentN}\end{equation}
and check that only the first two moments of the gaussian and ideal
responses are equal. The third central moment calculated for the ideal
response is $M_{3}^{c}=(1+3v_{1})\mu$ which neither coincides with
that of a normal distribution (it is simply zero), nor converges to
it with increasing $\mu$. Only the skew $s\equiv\frac{M_{3}^{c}}{\sigma^{\frac{3}{2}}}$,
which is a measure of the distribution asymmetry, indeed converges
to zero slowly enough as $\frac{1+3v_{1}}{(1+v_{1})^{\frac{3}{2}}}\frac{1}{\sqrt{\mu}}$.

Although the normal approximation of the scintillation line shape
is quite common \cite{Breitenberger}, there are situations in which
its use leads to systematic errors in the parameter definition. Two
examples will be considered below (see section \ref{sec:Two-examples}).
In order to resolve this problem, a better approximation of an ideal
scintillation shape is needed.

\section{The generalized gamma distribution as a limiting case for the ideal
response}

We will search for a function with the following properties:

\begin{enumerate}
\item the function converges to a normal distribution for $\mu\rightarrow\infty$;
\item it has the mean value and variance coinciding with that of the ideal
scintillator response;
\item it approximates the ideal scintillator response better than a conventional
normal distribution;
\item it is asymmetric with a skew decreasing as $\frac{1}{\sqrt{\mu}}$,
and gives a better approximation of the distribution tail.
\end{enumerate}
In literature the successful usage of the 2-parameter gamma- distribution
to approximate the output pulse height spectra of scintillation detectors
is reported, with better results in comparison with a normal approximation
\cite{Gamma},\cite{Stokey83}. We were not able to get a good agreement
with the response function of an ideal detector using the above- mentioned
distribution, so we have chosen a power transformed gamma distribution
(also known as generalized gamma distribution) as a candidate:

\begin{equation}
f(x;m,\alpha,\beta)=\frac{m}{\Gamma(\alpha)}\beta^{m\alpha}x^{m\alpha-1}e^{-(\beta x)^{m}}.\label{GenGamma}\end{equation}

The distribution describes a variety of well-known 1 and 2-parameter
probability laws as special cases; more details regarding the distribution
properties can be found in \cite{Hegyi}. A physical basis for the
generalized gamma distribution has been discussed by Lienhard and
Meyer in \cite{Lienhard}.

We start by fitting the ideal scintillator response for different
$\mu$ values using (\ref{GenGamma}) with 3 free parameters. It has
been discovered that over a wide region of $\mu$ the value of parameter
$m$ is close to 2, thus we fix it at this value and use the following
distribution as an approximation of the ideal shape response (redefining
$\beta^{2}$ from (\ref{GenGamma}) as $\beta$):

\begin{equation}
g(q;\alpha,\beta)=2\beta^{\alpha}\Gamma^{-1}(\alpha)q^{2\alpha-1}e^{-\beta q^{2}},\label{eq:Gamma}\end{equation}

with parameters $\alpha$ and $\beta$ providing equality of the mean
value and variance of (\ref{eq:Gamma}) to the corresponding values
of the ideal scintillation response. It is easy to check that the
moment of order $n$ of the distribution (\ref{eq:Gamma}) is:

\[
M_{n}=\beta^{-\frac{n}{2}}\frac{\Gamma(\alpha+\frac{n}{2})}{\Gamma(\alpha)}.\]

The parameters $\alpha$ and $\beta$ can be defined from the system
of equations:

\begin{equation}
\left\{ \begin{array}{c}
\overline{q}\equiv\mu=\frac{\Gamma(\alpha+\frac{1}{2})}{\Gamma(\alpha)}\beta^{-\frac{1}{2}}\\
\overline{q^{2}}\equiv\mu^{2}+\sigma^{2}=\frac{\alpha}{\beta}\end{array}\right.\label{eq:System}\end{equation}

A recipe for the approximate solution of the system is given in Appendix
A. An alternative way of calculating the parameters $\alpha$ and
$\beta$ based on the equality of the first two even moments of (\ref{eq:Gamma})
to the corresponding values of the ideal scintillation response, is
presented in Appendix B. 

It is important to stress that a special case $m=2$ is found in many
physical applications: in hydrology it is known either as hydrograh
distribution \cite{Lienhard}, or in countries where the Russian hydrology
school has become more familiar, as the Kritskiy- Menkel distribution
\cite{Kritskij}; in radio-engineering variants of the generalized
gamma-distribution are widely used to describe radio waves propagation
in fading environment (Nakagami distribution \cite{Nakagami}); some
further examples can be found in \cite{Schenzle}%
\footnote{In \cite{Hegyi} the case $m=2$ is called Stratonovich distribution.
We were unable to find the corresponding reference in literature.%
}.

\begin{figure}
\begin{centering}\includegraphics[width=0.35\paperwidth,height=0.35\paperwidth]{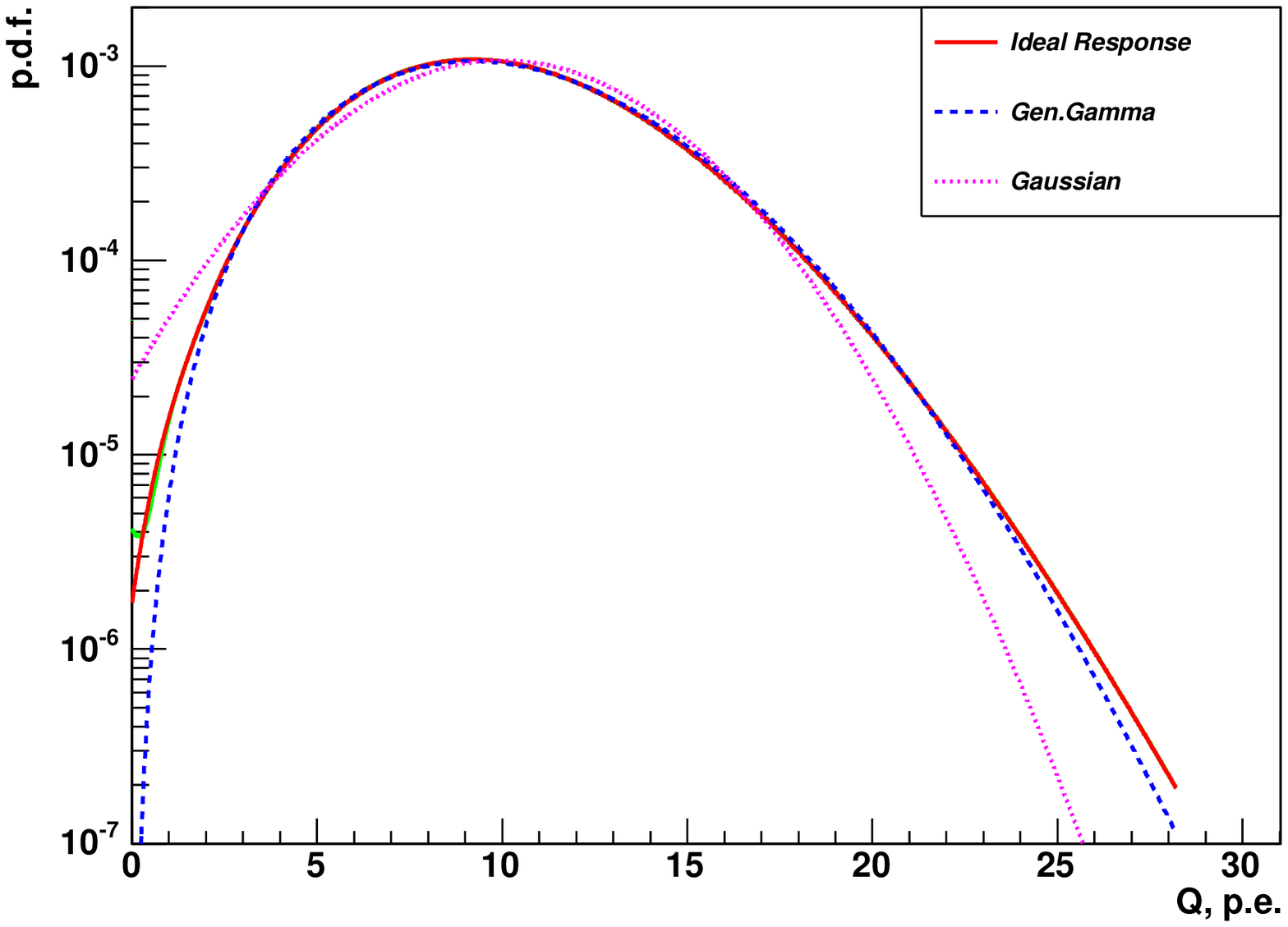}\includegraphics[width=0.35\paperwidth,height=0.35\paperwidth]{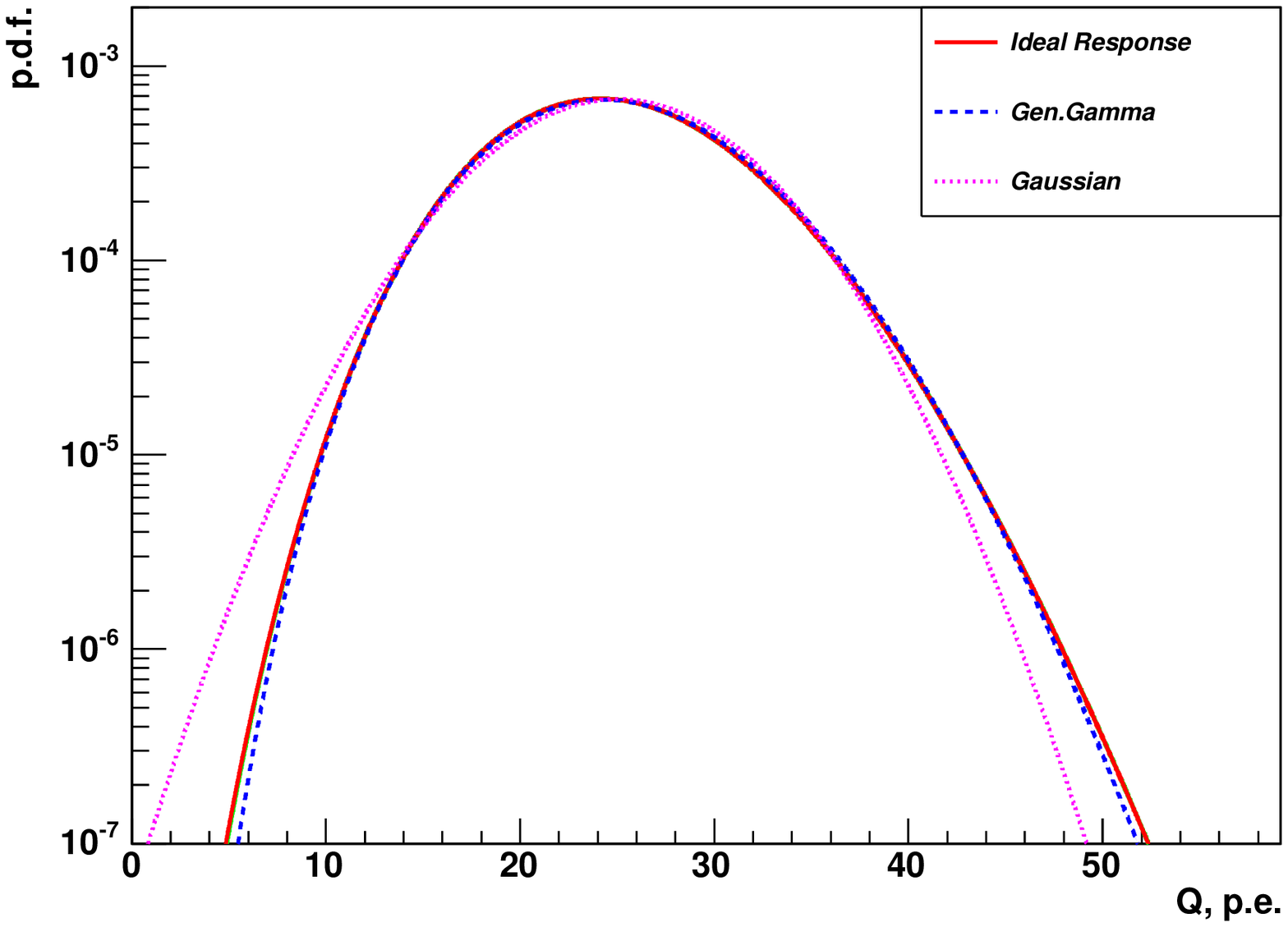}\par\end{centering}

\begin{centering}\includegraphics[width=0.35\paperwidth,height=0.35\paperwidth]{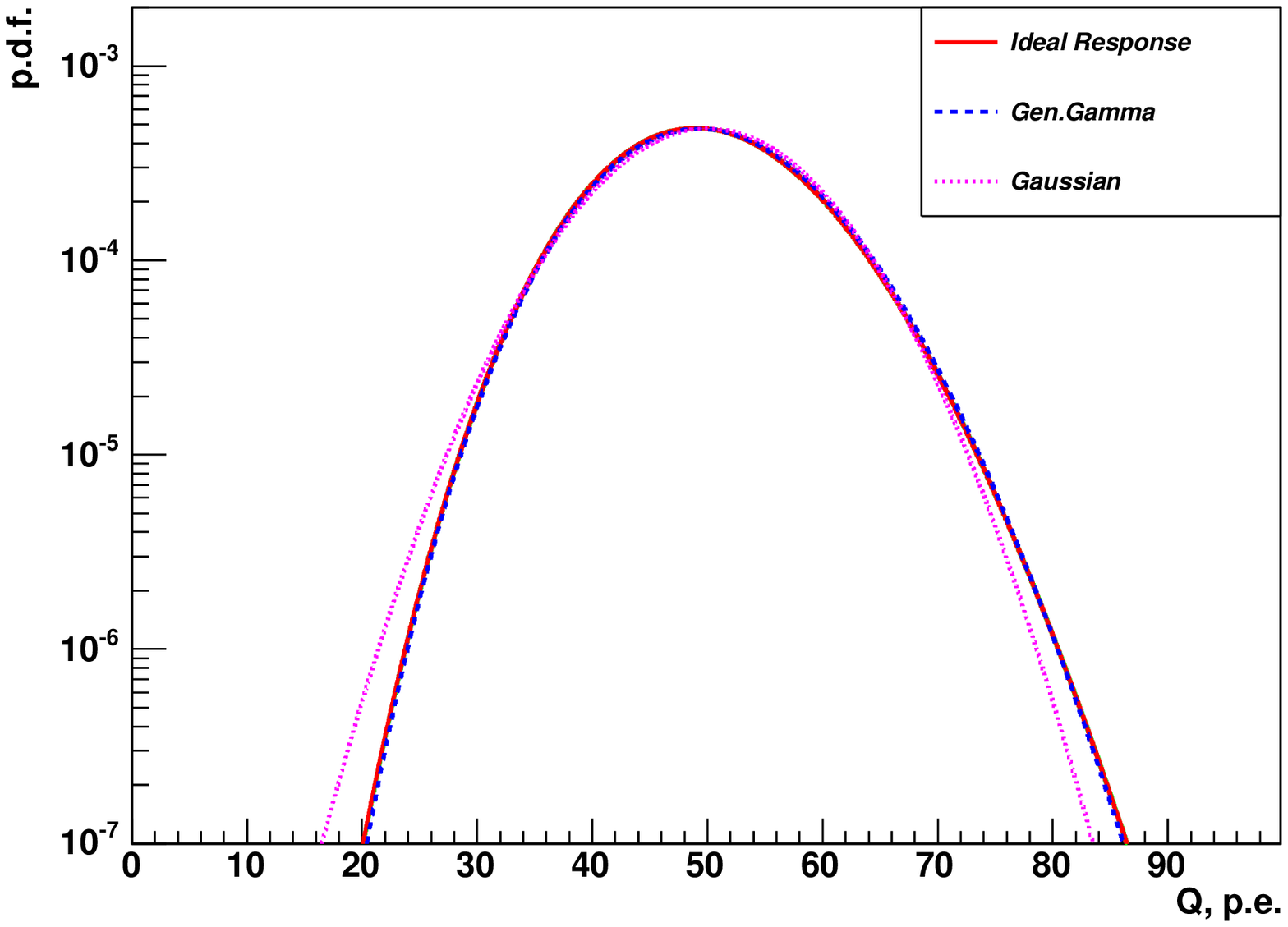}\includegraphics[width=0.35\paperwidth,height=0.35\paperwidth]{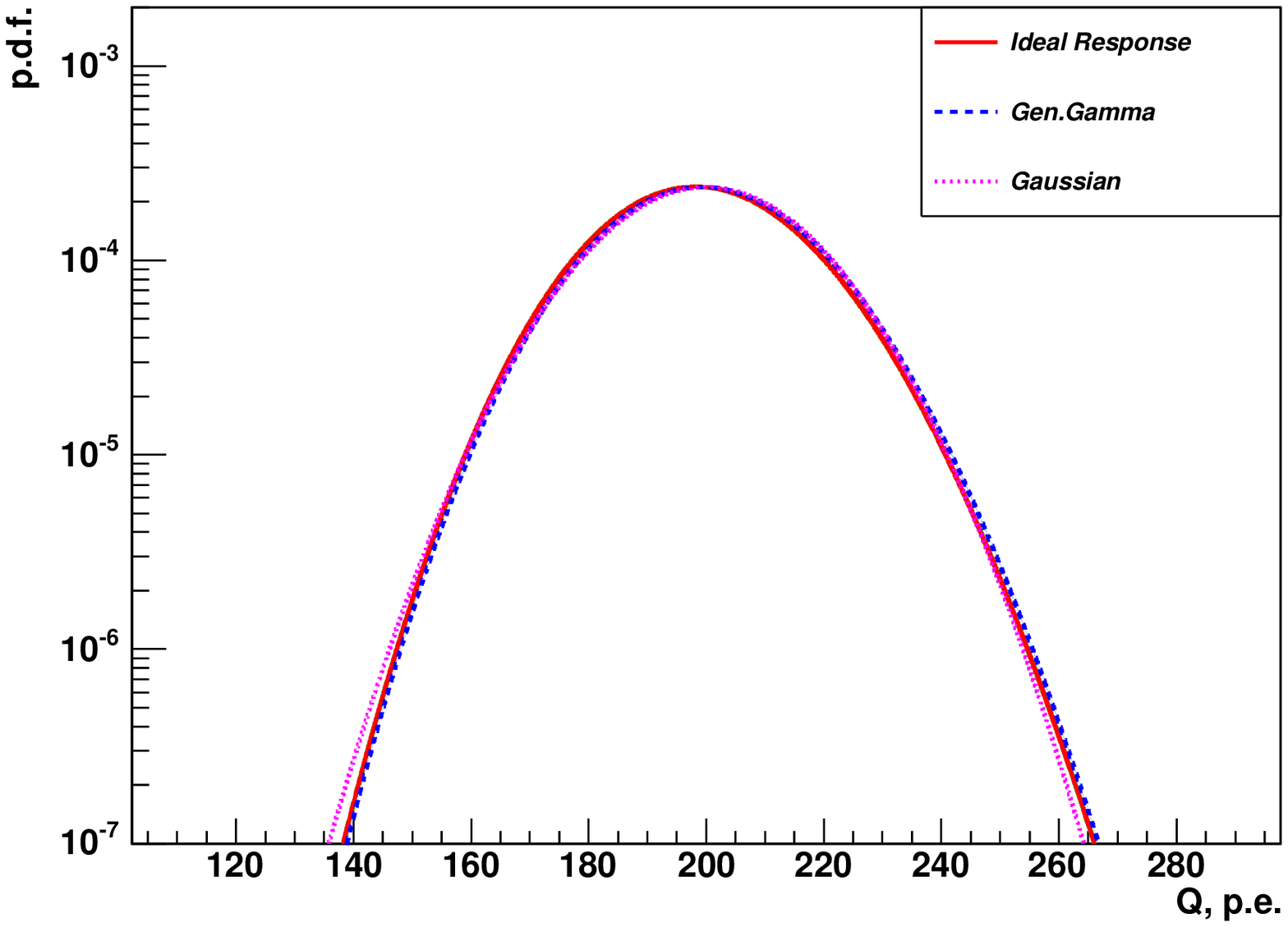}\par\end{centering}

\caption{\label{FigComparison}Comparison of the ideal scintillation response
with the gaussian and the model by means of a generalized gamma distribution
for $\mu=$10, 20, 50 and 200 p.e. Responses obtained by using the
realistic s.e.r. function (see Fig.\ref{figure:ser}) are not distinguishable
from the ideal scintillation response in all the above plots.}
\end{figure}

\begin{figure}
\begin{centering}\includegraphics[width=0.5\paperwidth,height=0.4\paperwidth]{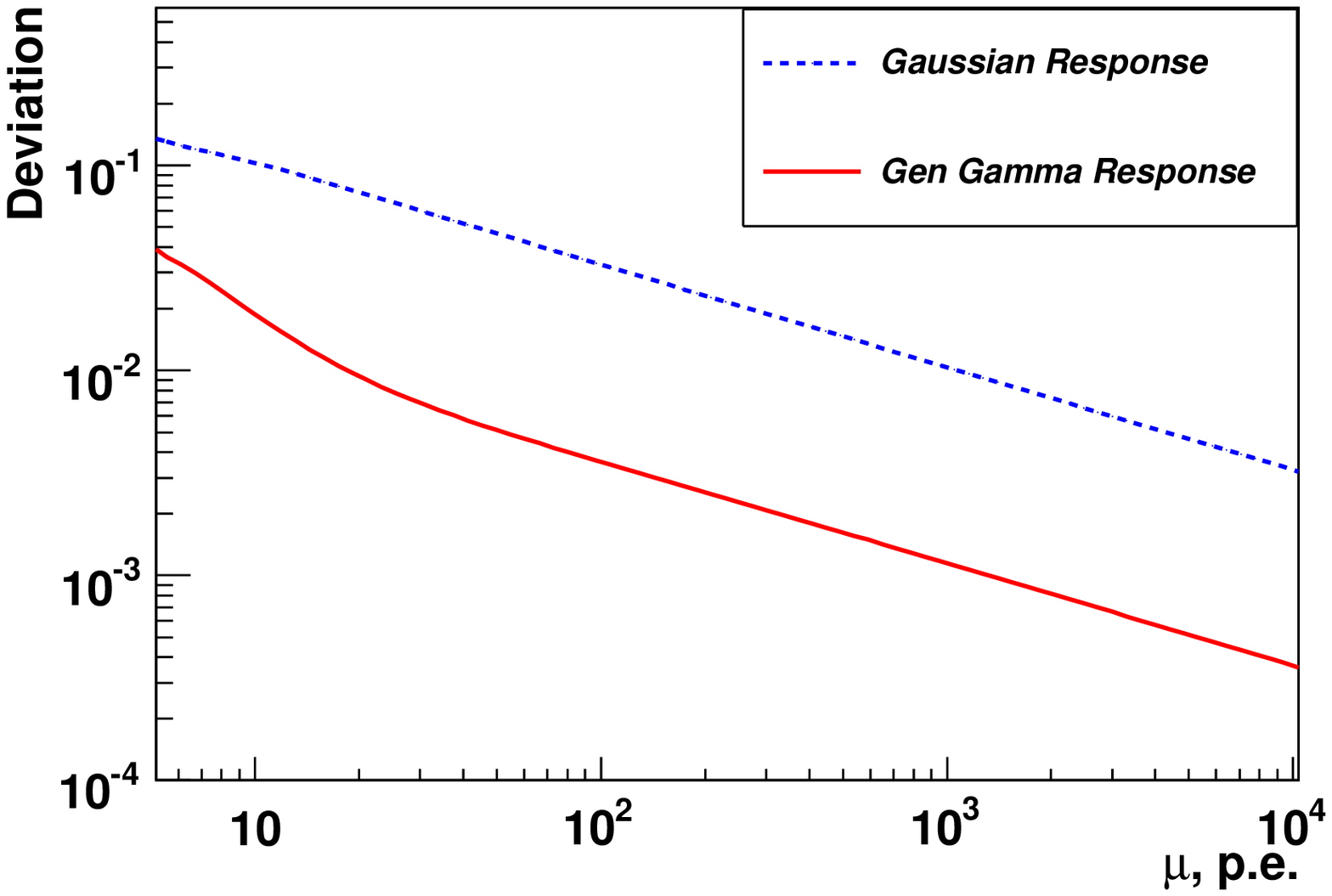}\par\end{centering}

\caption{\label{Modulus}The deviation of the response constructed by using
the generalized gamma function from the ideal one is an order of magnitude
lower than that for the corresponding gaussian. The deviation was
calculated by means of (\ref{Deviation}).}
\end{figure}

In the limit $\alpha\rightarrow\infty$ the distribution $g(q)$ converges
to a normal distribution \cite{Haken}, the condition 2 is satisfied
automatically, conditions 3 and 4 have been checked numerically in
a wide range of $\mu$ values. As it can be seen in Fig.\ref{FigComparison}
the generalized gamma distribution approximates the ideal response
better than a gaussian. Fig.\ref{Modulus} presents results of numerical
calculations of the deviation of the gaussian (with the mean value
and variance that of an ideal response) and the shape obtained with
(\ref{eq:Gamma}) from the ideal response calculated as:

\begin{equation}
\int_{\mu-5\sigma}^{\mu+5\sigma}|g(q)-f(q)|dq,\label{Deviation}\end{equation}
and has a simple mathematical interpretation. In Fig.\ref{Modulus}
one can see that the deviation of the generalized gamma-distribution
from the ideal one calculated by using (\ref{Deviation}) is an order
of magnitude lower than that in the gaussian distribution case.

The quality of the fit in the tail has been checked by calculating
the integral in the region $[\mu+2\sigma;\infty]$ for the ideal and
generalized gamma- distributions. The integral of the gaussian in
this region is constant defined by the complementary errors function:
$0.5\mathrm{\mathrm{{erfc}}}(\sqrt{2})$. The cumulative distribution
corresponding to the density (\ref{eq:Gamma}) is:

\begin{equation}
G(x)\equiv\int_{0}^{x}g(x)dx=\gamma(\alpha,\beta x^{2}),\label{GammaCum}\end{equation}

where $\gamma(\alpha,x)$ is the normalized incomplete gamma function.
The integral in the tail is $1-G(\mu+2\sigma)$.

\begin{figure}
\begin{centering}\includegraphics[width=0.5\paperwidth,height=0.4\paperwidth]{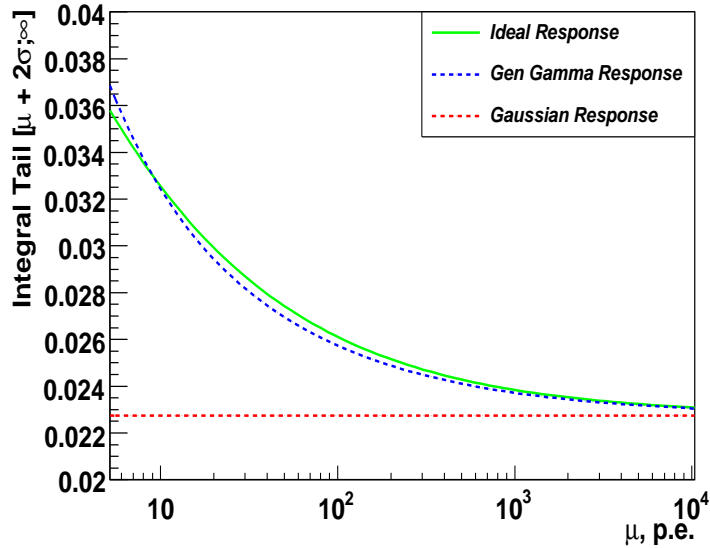}\par\end{centering}

\caption{\label{Tails}The ideal scintillation response tail is reproduced
very well for $\mu\gtrsim8$ p.e. The corresponding gaussian response
tail does not depend on $\mu$ and is defined by $\frac{1}{2}\mathrm{{erfc}}(\sqrt{2})$.}
\end{figure}

Integral in the tail for the ideal response was calculated by using
the original definition (\ref{PMT_pdf}):

\[
t=\sum_{n=N_{min}}^{n=N_{max}}P(n)\frac{1}{2}\mathrm{{erfc}}\left(\frac{2\sigma}{\sqrt{2v_{1}n}}\right),\]
with $N_{min}=max([\mu-2\sigma],0)$ and $N_{max}=\mu+5\sigma$. The
results are presented in Fig.\ref{Tails}. One can see that the gamma
distribution gives a better approximation of the distribution tail
than the gaussian one.

The most probable value of distribution (\ref{eq:Gamma}) corresponds
to $\hat{q}=\sqrt{\frac{1}{\beta}(\alpha-\frac{1}{2})}$ \cite{Schenzle},
it can be seen that $\hat{q}$ is shifted to the left from the mean
value $\mu$ by $\simeq\frac{1+v_{1}}{2}$.

\section{\label{sec:Two-examples}Two examples}

The precision of the description of the spectra of a real scintillation
detector with respect to different approximations of the response
function has been verified by using both the real data of the Counting
Test Facility (CTF, \cite{CTF}) of the Borexino detector$\;$\cite{BORgen},
and the data obtained with the Monte Carlo model of the CTF detector.
In the present article we consider only the MC data, the results of
comparison of the theoretical model with the real CTF data will be
presented by the Borexino collaboration.

The large volume liquid scintillator detector CTF is a prototype of
the solar neutrino detector Borexino. The CTF was used to develop
the methods of deep purification of the liquid scintillator and water
from the natural radioactive impurities. The CTF consists of 3.7 tones
of liquid scintillator on the base of pseudocumene (C$_{9}$H$_{12}$),
contained in a transparent spherical inner vessel with a radius of
1 m, and viewed by 100 photomultipliers (PMTs) mounted on an open
spherical steel support structure. The PMTs are equipped with light
concentrator cones to increase the light collection efficiency; the
total geometrical coverage of the system is $21\%$. The radius of
the sphere passing through the opening of the light cones is 2.73
m. The entire detector is placed inside a cylindrical tank with water,
which provides shielding against external gammas. On the bottom of
the tank another 16 PMTs are mounted to identify cosmic muons by their
Cherenkov light produced in the water. The detailed description of
the CTF detector can be found in \cite{CTF}. The CTF has been in
operation since 1993. At present it is in its third data-taking campaign
(CTF3) with the main goal of tuning the purification strategy for
the Borexino detector. The data collected with an upgraded version
of the CTF were used by Borexino collaboration in order to search
for a number of possible manifestations of non-standard physics, a
review of experimental results can be found in \cite{REVIEW}.

The Monte Carlo model of the CTF detector was developed on the basis
of EGS-4 code \cite{EGS4} to check the validity of the background
interpretation. It accounts for the dependence of the light yield
on the energy (ionization quenching) and on the position where energy
was deposited inside the detector. The model has been calibrated with
the CTF data and describes the CTF experimental spectra with a satisfactory
precision. For the purposes of the present work, the model of the
detector response was changed to take into account the deviations
of the response function from the normal one (the standard program
uses the normal approximation of the response function).

\subsection{Monoenergetic line}

The detector response to the monoenergetic particle has been modeled
with the MC method. The particle energy was chosen in order to provide
the number of registered photoelectrons, $\mu=150$ p.e. The number
is big enough to ensure good approximation with a gaussian shape.
Indeed, the processing of the CTF data by using this approximation
was successfully applied even for lower values of the mean registered
charge \cite{BorexEM}.

The response of the detector was generated in the following way. First,
the mean number $\mu_{0}$ of p.e. registered at one PMT was defined
as $\mu_{0}=\mu/N_{PMT}$, where $N_{PMT}$ is the total number of
the PMTs in the detector. Then in each event for each PMT the Poisson-
distributed number $K$ of registered p.e. was generated, and, finally,
the registered anode charge was simulated using the gaussian approximation
of the PMT signal with mean $\mu=K$ and variance $\sigma_{\mu}^{2}=v_{1}K$.
The response of the detector is the sum of signals over all PMTs of
the detector. $N=10^{6}$ events were simulated.

\begin{figure}
\begin{centering}\includegraphics[width=0.35\paperwidth,height=0.35\paperwidth]{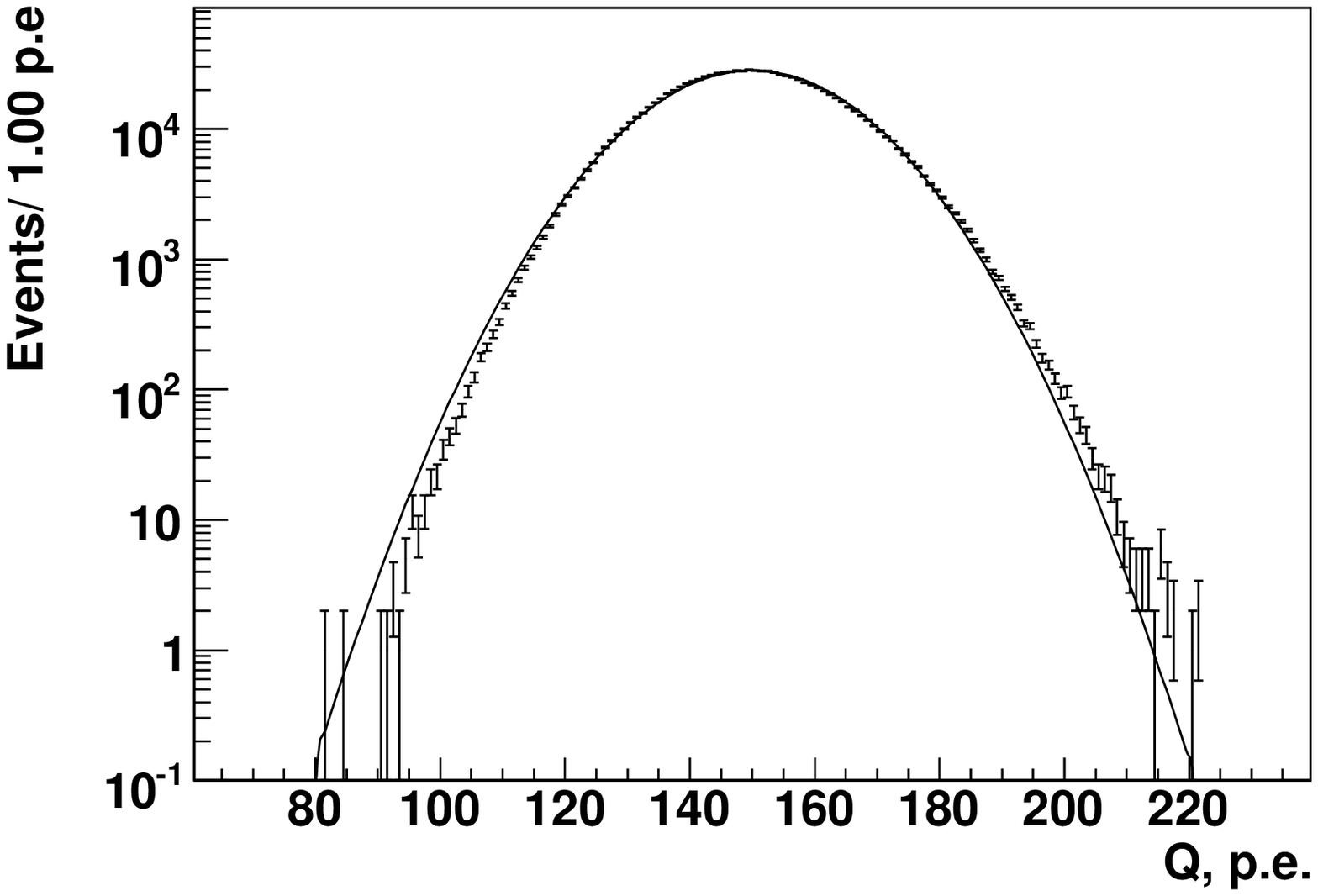}\includegraphics[width=0.35\paperwidth,height=0.35\paperwidth]{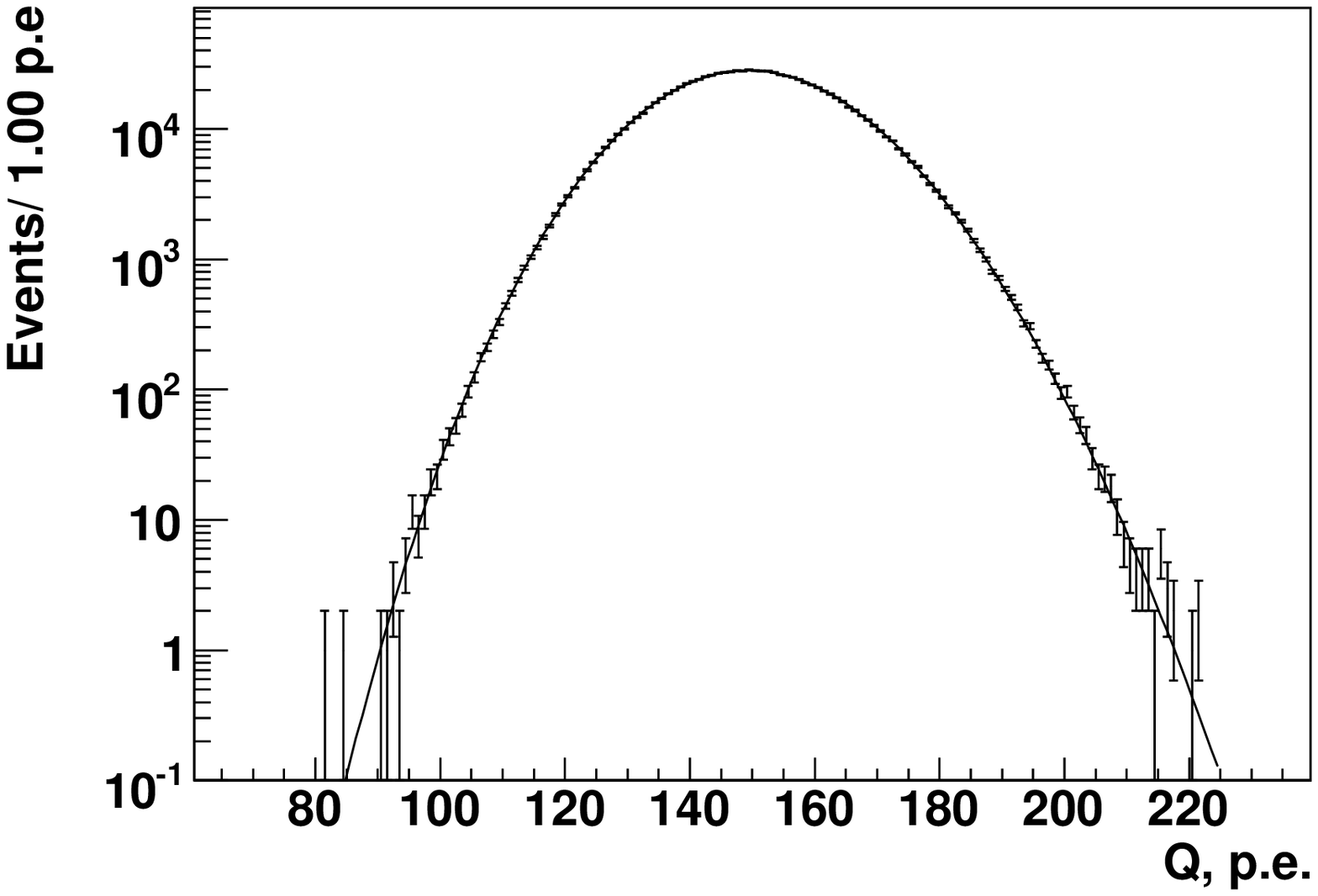}\par\end{centering}

\caption{\label{MC_Alpha_gauss}Comparison of the MC generated monoenergetic
response fit using the normal (left) and generalized gamma (right)
distributions. To the non-critical eye the both fits are comparable
in the region $\mu\pm2\sigma$, however, the deviations in the tail
for the gaussian distribution are evident. The $\chi^{2}=111.6$ value
for the generalized gamma distribution is close to the number of the
degrees of freedom (n.d.f.=116) while for the normal distribution
the $\chi^{2}/n.d.f=1883/116$ excludes the normal-distribution hypothesis.}
\end{figure}

The MC data were fit with the gaussian response function and with
the response function based on the generalized gamma- distribution.
The results of the fit are presented in Table \ref{MonoETable} and
Fig.\ref{MC_Alpha_gauss}. The mean values and the normalization are
reproduced well for the gaussian and generalized gamma line shapes;
the difference in variances is within the statistical precision of
the method. The $\chi^{2}$ value for the gaussian case excludes the
hypothesis of the normal line shape; in the case of the non-gaussian
shape we have a good match of the data with the model ($\chi^{2}/n.d.f.=111.6/116$,
the number of degrees of freedom (n.d.f.) here is the number of bins
used in the fit with the number of free parameters subtracted). We
have found no difference when applying method A or B (see Appendix
A and B) to estimate of parameters of the non-gaussian line shape. 

As it is noted above, Prescott in \cite{Prescott63} obtained a precise
line shape for the case of an exponential single photoelectron response
$f_{1}(x)=\frac{1}{a}e^{-\frac{x}{a}}$, $x\geq0$, it reads:

\begin{equation}
f(x)=\frac{1}{a}\sqrt{\mu}e^{-\mu}\left(\frac{x}{a}\right)^{-\frac{1}{2}}e^{-\frac{x}{a}}I_{1}(2\sqrt{\mu\frac{x}{a}}),\label{Prescott0}\end{equation}
where $I_{1}$ is a modified Bessel function of the first kind for
an imaginary argument. 

The slope of an exponential distribution coincides with its mean value,
i.e. $q_{1}=a$. The variance of the single electron exponential response
doesn't depend on parameter $a$ and is $v_{1}^{exp}=2$. It is clear
that formula (\ref{Prescott0}) can't be directly applied to fit the
real scintillation shape. The way to solve this problem was pointed
out in \cite{Prescott63}: it is enough to treat $a=\frac{\sigma_{Q}^{2}}{2\mu}$
as a scale parameter, the variance in this case will scale as $\sqrt{a}$
and the mean value as $a$. In order to preserve the mean value and
variance in the original scale, we multiply $\mu$ by a scale parameter
$s=\frac{2\mu}{\sigma_{Q}^{2}}=\frac{2}{1+v_{1}}$, and as before
set $q_{1}=1$:

\begin{equation}
f(x)=s\sqrt{\mu s}e^{-\mu s}(xs)^{-\frac{1}{2}}e^{-xs}I_{1}(2s\sqrt{\mu x}).\label{Prescott}\end{equation}

Now formula (\ref{Prescott}) can be used to fit the scintillation
line, the results are presented in Table \ref{MonoETable}. Comparing
the $\chi^{2}$ values one can see that the quality of the fit with
Prescott formula is worse than in the case of the fit with the generalized
gamma function, but much better than in the case of the fit with the
normal distribution. The quantitative comparison of the models can
be performed using Fischer's F-distribution as a significance test:
$\frac{\chi_{2}^{2}}{\chi_{1}^{2}}=F(\alpha,\nu,\nu)$, where $\nu$
is a number of the degrees of freedom and $\alpha$ is a confidence
level \cite{Wolberg}. Solving equation $F(\alpha,116,116)=1883/111.6$
with respect to $\alpha$ one can exclude the gaussian shape with
a c.l. more than 99.99$\%$. The scintillation line shape is described
better by Prescott's formula (as can be seen from the comparison of
$\chi^{2}$ values in Table \ref{MonoETable}) and the exclusion c.l.
is smaller, but Prescott's model fails to describe the data with high
precision as the generalized gaussian distribution does. 

The obtained results have demonstrated very weak sensitivity of the
real line shape to the shape of the s.e.r., so one can choose any
convenient s.e.r. shape in order to invert formula (\ref{FTransform2}).

\begin{table}
\begin{centering}\begin{tabular}{|c|c|c|c|c|}
\hline 
&
$\mu$&
$\sigma_{Q}$&
Norm ($\times10^{6}$)&
$\chi^{2}/n.d.f.$\tabularnewline
\hline
\hline 
MC input&
150.00&
14.18&
1.000&
\tabularnewline
\hline 
Gauss&
150.01$\pm$0.05&
14.19$\pm$0.03&
1.000$\pm$0.001&
1883/116\tabularnewline
\hline 
Gen.gamma&
150.02$\pm$0.05&
14.17$\pm$0.03&
1.000$\pm$0.001&
111.6/116\tabularnewline
\hline 
Prescott&
149.52$\pm$0.05&
14.19$\pm$0.03&
1.000$\pm$0.001&
329.0/116\tabularnewline
\hline
\end{tabular}\par\end{centering}

\caption{\label{MonoETable}Characteristics of three different fits of the
monoenergetic line.}
\end{table}

\subsection{$^{14}C$ beta spectrum: MC model of the experimental data}

The major part of the background in the ultra-pure CTF in the energy
region up to 200 keV is induced by $\beta$-activity of $^{14}C$
\cite{CTF_C14}, which is present in the organic liquid scintillator
at the level of $10^{-18}$ g/g. The $\beta$-decay of $^{14}C$ is
an allowed ground-state to ground-state ($0^{+}\rightarrow1^{+}$)
Gamow-Teller transition with an endpoint energy of $E_{0}=156$$\;$keV
and half life of 5730 years. The end-point of the decay is used in
CTF to establish the energy scale, thus the precision of the modeling
of $^{14}$C spectrum defines the precision of the energy scale calibration.

The beta energy spectrum with a massless neutrino can be written in
the following form \cite{Morita}:

\begin{equation}
dN(E)\sim F(Z,E)C(E)pE(Q-E)^{2}dE\label{BetaSpectrum}\end{equation}

where 

$E$ and $p$ \textbf{}are the total electron energy and momentum; 

$F(E,Z)$ is the Fermi function with correction of screening caused
by atomic electrons; 

\textbf{$C(E)$} contains departures from the allowed shape. 

For $F(E,Z)$ we have used the function from \cite{Simpson} which
agrees with tabulated values of the relativistic calculation \cite{RElativistic}.
A screening correction has been made by Rose's method \cite{Rose}
with screening potential $V_{0}=495$$\;$eV. The $^{14}C$ spectrum
shape factor can be parametrized as $C(E)=1+\alpha E$ (see \cite{Kuzminov}
for more details), the value of the parameter $\alpha$ was fixed
at the value $\alpha=-0.7$ MeV$^{-1}$. 

The deviations of the light yield from the linear law have been taken
into account by using the ionization deficit function $f(k_{B},E)$,
where $k_{B}$ is Birks' constant \cite{birks}. To calculate the
ionization quenching effect for the scintillator on the base of pseudocumene,
we used the KB program from the CPC library \cite{KB}. The value
of the ionization quenching parameter $k_{B}=0.017$ cm$^{-1}$MeV$^{-1}$
was fixed at the value found by independent experiments. The radial
dependence of the mean registered charge on the point of interaction
inside the detector has been accounted for with the $f_{R}(r)$ function,
obtained from the experimental data (see \cite{Resolutions}). For
convenience the value of the $f_{R}$ function at the detector's center
was assumed to be the unity, $f_{R}(0)=1$.

The response of the detector for an event of $^{14}$C decay was generated
in the following way. First, the event energy $E$ was generated according
to the spectrum (\ref{BetaSpectrum}), and the position of the event
was generated in assumption of uniform distribution of $^{14}$C decay
events in the detector volume. Then the mean number of p.e. has been
defined, registered for an event of energy $E$ occurring at distance
$r$ from the detector center, taking into account detector's non-uniformity
and non- proportionality of the light yield on the energy:

\[
Q(E,r)=A\cdot E\cdot f_{R}(r)\cdot f(k_{B},E),\]

where \textbf{$A$} is the scintillator specific light yield measured
in photoelectrons per MeV.

Then in each event for each PMT the mean value of registered number
of p.e. has been defined, and the registered p.e. number $K$ was
generated according to the corresponding Poisson distribution. Finally,
the registered anode charge was simulated by using a gaussian approximation
of the PMT signal with mean $\mu=K$ and variance $\sigma_{\mu}^{2}=v_{1}K$.
The response of the detector is the sum of the signals over all PMTs
of the detector. $N=5\times10^{7}$ event were simulated, that corresponds
approximately to 3 years of continuous data taking with the CTF detector.

The exponential underlying background has been added to the $^{14}$C
$\beta$-spectrum to simulate the realistic situation. We have taken
the parameters of the exponential observed in the CTF detector. This
background is mainly due to the external $\gamma$'s from decays of
elements from $^{238}$U and $^{232}$Th chains in the water shield.

\subsection{\label{Carbon14subsection}$^{14}C$ beta spectrum: fitting MC data
with model function}

The real detector response to uniformly distributed events is not
spatially uniform. To take into account the additional pulse height
variance we exploit formula \cite{Resolutions}:

\begin{equation}
\sigma_{Q}^{2}=(1+\overline{v_{1}})Q+v_{T}Q^{2},\label{RES}\end{equation}

where 

\begin{lyxlist}{00.00.0000}
\item [{$Q=A\cdot E\cdot f(k_{B},E)\cdot\overline{f_{R}}$}] is the mean
total registered charge for the events of the energy E uniformly distributed
over the detector volume. $\overline{f_{R}}$ is the mean value of
the $f_{R}(r)$ function over the detector volume;
\item [{\textbf{$\overline{v_{1}}=\frac{1}{N_{PMT}}\sum_{i=1}^{N_{PMT}}s_{i}v_{1_{i}}$}}] is
the relative variance of the PMT single photoelectron charge spectrum
($v_{1_{i}}$) averaged over all PMTs of the detector ($N_{PM}$ in
total) taking into account the i-th PMT relative sensitivity $s_{i}$.
For the CTF detector this parameter has been defined with a high precision
during acceptance tests \cite{2200PMTs} and turns out to be $\overline{v_{1}}=0.34$;
\item [{\textbf{$A$}}] is the scintillator specific light yield measured
in photoelectrons per MeV;
\item [{$v_{T}$}] is the relative variance of the photon transfer efficiency,
mainly due to the spatial non-uniformity of the detector. Among other
additional contributions there is the intrinsic scintillator line
width, the precision of the detector calibration, the precision of
zero signal definition, etc. There is now need to keep these additive
parameters apart, so in the model we have left the only parameter.
In the MC modeling these additional contributions were set to zero,
but, nevertheless, parameter $v_{T}$ remained free, see discussion
below.
\end{lyxlist}
The MC spectrum was modeled with a sum of two components: (1) convolution
of the $^{14}C$ beta spectrum with the detector resolution function
with 3 free parameters: total normalization $N$, light yield $A$,
and additional variance $v_{T}$; (2) an additional exponential background
with 2 free parameters.

The final model function $S(Q)$ has 5 free parameters and is presented
as: 

\begin{equation}
S(Q)=N_{0}\int N(E(Q'))\frac{dE}{dQ}Res(Q,Q')dQ'+ExpBkg(Q),\label{Model}\end{equation}

where $Res(Q,Q')$ is the detector response function, and N(E) is
the $^{14}C$ beta- spectrum (\ref{BetaSpectrum}).

\begin{figure}
\begin{centering}\includegraphics[width=0.8\paperwidth,height=0.2\paperwidth]{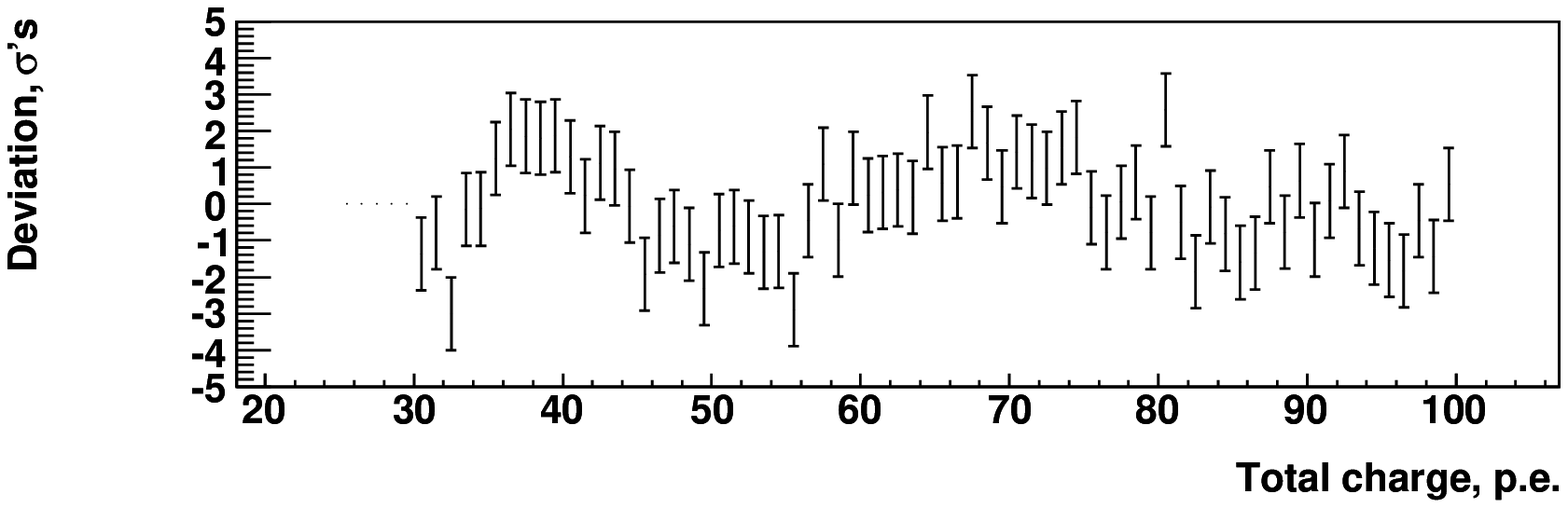}\par\end{centering}

\begin{centering}\includegraphics[width=0.8\paperwidth,height=0.2\paperwidth]{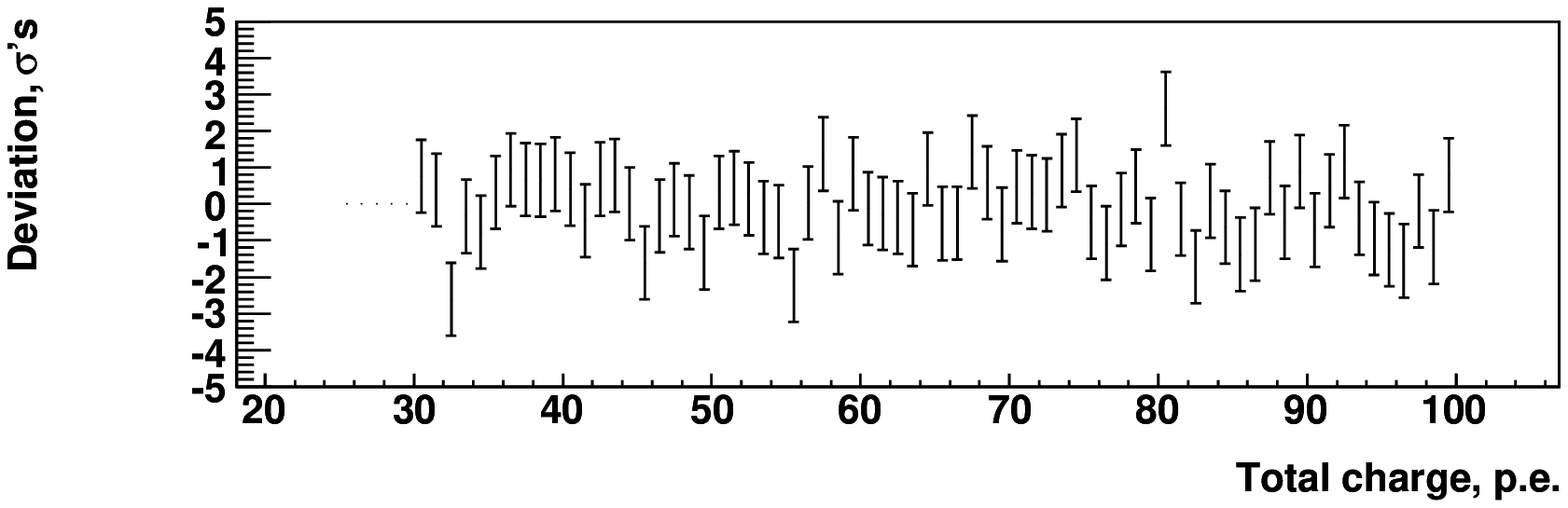}\par\end{centering}

\caption{\label{Figure:ChiSq}Residual of the fit of the data using the normal
and generalized gamma distributions (the region up to 100 p.e. is
shown). The residual of the fit with the normal distribution (upper
plot) has two fake peaks in the region of the $^{14}C$ tail. This
is a typical situation for the resolution function mismatch. The fit
of the same data with the generalized gamma function (lower plot)
has no pronounced artifacts in the region of the $^{14}C$ beta-spectrum
tail. }
\end{figure}

\begin{figure}
\begin{centering}\includegraphics[width=0.8\paperwidth,height=0.4\paperwidth]{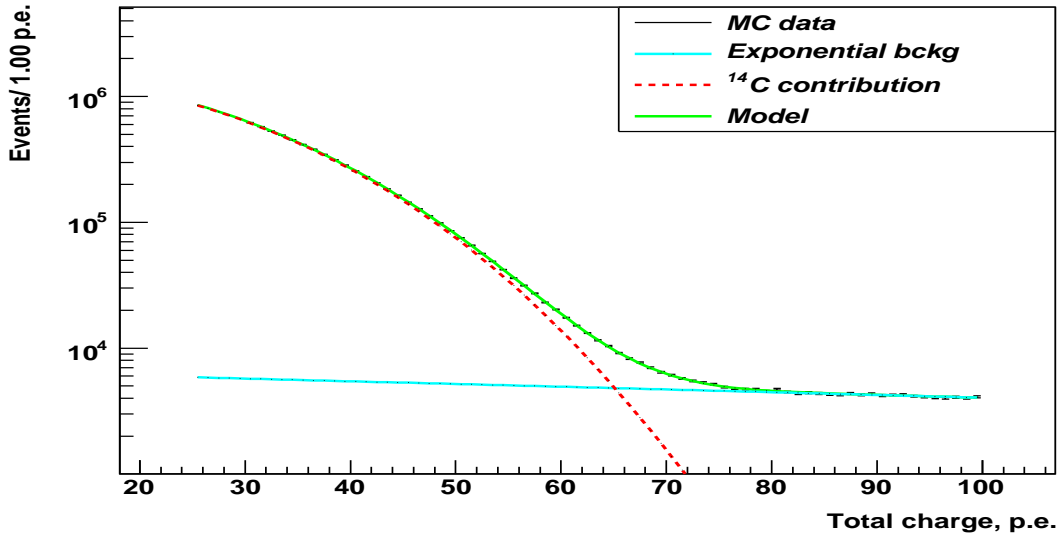}\par\end{centering}

\caption{\label{C14}Fit of MC $^{14}$C spectrum with a model function (only
the region up to 100 p.e. is shown). The fit region 30-250 p.e. corresponds
to 91-681 keV.}
\end{figure}

\begin{table}
\begin{centering}\begin{tabular}{|c|c|c|c|c|}
\hline 
&
$A$&
Norm ($\times10^{6}$)&
Slope&
$\chi^{2}/n.d.f.$\tabularnewline
\hline
\hline 
MC input&
391.8&
5.000&
100.0&
\tabularnewline
\hline 
Gauss&
387.8$\pm$0.3 (-13$\sigma$)&
5.174$\pm$0.010 (+17$\sigma$)&
99.2$\pm$0.5 ($-2\sigma$)&
279.7/214\tabularnewline
\hline 
Gen.Gamma&
394.0$\pm$0.3 (+7 $\sigma$)&
5.033$\pm$0.008 (+4$\sigma$)&
100.0$\pm$0.3 (0$\sigma$)&
211.3/214\tabularnewline
\hline
\end{tabular}\par\end{centering}

\caption{\label{C14Table}Parameters of the model fitting the CTF MC $^{14}$C
spectrum. Errors cited for each parameter are 68\% c.l. errors obtained
while studying the $\chi^{2}$-profile. The value in parenthesis near
every fitting parameter gives a deviation from the nominal value in
units of the standard deviation for the corresponding parameter.}
\end{table}

The results of the fit of the experimental data with the gaussian
and non-gaussian line shapes in 30-250 p.e. region, are presented
in Table \ref{C14Table}. Again, the $\chi^{2}$ is much better for
the non-gaussian line shape. The comparison of the models excludes
the gaussian shape on the c.l. of 98$\%$ (solution of $F(\alpha,\nu,\nu)=279.7/211.3$
with $\nu=214$ gives $\alpha=2\times10^{-2}$).

This time relatively big deviations in parameters have been found
when applying different resolution functions. The deviations for parameters
are bigger than statistically allowed, so it should be treated as
systematic errors. As it follows from Table \ref{C14Table}, the error
in the light yield definition for the case of the gaussian line shape
is $-1\%$, the error of the total normalization is $+3.5\%$. With
the generalized gamma function the error in light yield is smaller:
$+0.6\%$, the same error has the total normalization. 

It is not implicitly assumed that additional broadening of the scintillation
line shape ($v_{T}Q^{2}$) is distributed in the same way as the main
contribution $(1+\overline{v_{1}})Q$. The statement is not true in
general, especially for big $Q$ values where $v_{T}Q^{2}$ term can
dominate in the response. In our case the main term dominates, that
is confirmed by the quality of the fit, so the precise distribution
for the additional line broadening can be neglected. The price paid
for this simplification is the observed systematical deviations. 

When fitting the monoenergetic line from $\alpha-$decays of $^{214}$Po
without selecting the detector central region the quality of the fit
is much worse at the left side of the peak. In the case of $^{14}$C
spectrum these imperfections on the left side are covered due to the
fast decrease in the spectrum and the gaussian shape is justified.
On the right side the proper description of the scintillation line
tail is important because of the same fact of the fast decrease of
the spectrum. In the case of the monoenergetic line the true shape
of the distribution of the mean values over the detector volume, has
to be taken into account.

\section{Conclusions}

An approximation of the real line shape of the scintillation detector
with the generalized gamma distribution has been proposed. The approximation
describes the ideal scintillation line shape better than the widely
used normal distribution. Two parameters of the proposed function
are uniquely defined by the first two moments of the detector response
or by the first two even moments. The computational complexity of
the resolution function calculation is comparable to that of the normal
resolution.

It has been demonstrated that the ideal detector response to many
photoelectrons ($\mu\gtrsim8$) loose the sensitivity to the shape
of the single electron response of a photomultiplier and the only
important parameter is the s.e.r. relative variance. In analytical
calculations any convenient function can be used instead of a real
s.e.r.

While for the relatively \char`\"{}flat\char`\"{} experimental spectra
one can hardly expect the enhancement of the overall quality of the
fit, in the case of the fast-varying distributions, such as tails
of the $\beta-$spectrum, the use of the proposed resolution function
allows one to exclude the artifacts associated with resolution mismatch,
and avoid systematics errors as demonstrated by the example with the
$^{14}$C spectrum fit.

\section*{Acknowledgments}

I am very grateful to Ferenc Dalnoki-Veress and Svetlana Chubakova
for the careful reading of the manuscript and useful discussions.

\section*{Appendix A }

An approximate solution of system (\ref{eq:System}) can be obtained
using the following expansion \cite{Graham1994}:

\begin{equation}
\frac{\Gamma(\alpha+\frac{1}{2})}{\Gamma(\alpha)}=\sqrt{\alpha}\left(1-\frac{1}{8\alpha}+\frac{1}{128\alpha^{2}}+\frac{5}{1024\alpha^{3}}-\frac{21}{32768\alpha^{4}}+...\right)\label{eq:Expansion}\end{equation}

For big $\mu$ the expansion converges fast because of $\alpha\sim\mu$.
Taking three first terms and substituting $\beta$ in the first equation,
we obtain a simple quadratic equation 

\[
f(\alpha)\equiv1-\frac{1}{8\alpha}+\frac{1}{128\alpha^{2}}=\frac{\mu}{\sqrt{\mu^{2}+\sigma^{2}}}\]

with the only positive root:

\begin{equation}
\alpha_{0}=\frac{1+\sqrt{\frac{2\mu}{\sqrt{\mu^{2}+\sigma^{2}}}-1}}{16(1-\frac{\mu}{\sqrt{\mu^{2}+\sigma^{2}}})},\label{eq:Alpha0}\end{equation}

which gives the solution with a relative precision of $\sim10^{-3}$
for $\mu>10$. A more accurate solution can be obtained by using more
terms from the expansion (\ref{eq:Expansion}). Assuming that more
accurate solution has a form $\alpha=\alpha_{0}+\Delta\alpha$ and
developing $f(\alpha)$ and two remaining terms from (\ref{eq:Expansion})
into a Tailor series keeping only a linear term with respect to $\Delta\alpha$,
we obtain a linear equation for $\Delta\alpha$ with the following
solution:

\begin{equation}
\Delta\alpha=\frac{\frac{21}{32}-5\alpha_{0}}{128\alpha_{0}^{2}-16\alpha_{0}-15+\frac{21}{8\alpha_{0}}}.\label{eq:DeltaAlpha}\end{equation}

Equation (\ref{eq:DeltaAlpha}) gives the relative precision of the
parameter estimation of $\lesssim10^{-4}$ at $\mu=20$, at $\mu=100$
it is $\simeq10^{-7}$.

\section*{Appendix B}

In radio-engineering the generalized gamma-distribution variant are
widely used to describe radio waves propagation in fading environment.
One of the most popular is the m-distribution proposed by Nakagami
\cite{Nakagami} in the functional form

\[
p(R)=\frac{2m^{m}R^{2m-1}}{\Gamma(m)\Omega^{m}}e^{-\frac{m}{\Omega}R^{2}},\]

where $\Omega=\overline{R^{2}}$, and $m$ is the inverse of the relative
variance of $R^{2}$. The advantages of this equation are simple rules
to calculate the parameters. 

In fact, for the even moments of (\ref{eq:Gamma}) the system of two
equations for $\alpha$ and $\beta$ will not contain gamma- functions.
Using the parameters $\alpha$ and $\beta$ we can write the second
and the fourth moments:

\begin{equation}
\left\{ \begin{array}{c}
\overline{q^{2}}=\frac{\alpha}{\beta}\\
\overline{q^{4}}=\beta^{-2}\frac{\Gamma(2+\alpha)}{\Gamma(\alpha)}=\overline{q^{2}}\cdot(\overline{q^{2}}+\frac{\overline{q^{2}}}{\alpha})\end{array}\right..\label{eq:System2}\end{equation}

The solution of this system is

\begin{equation}
\left\{ \begin{array}{c}
\alpha=\frac{\left(\overline{q^{2}}\right)^{2}}{\overline{q^{4}}-\left(\overline{q^{2}}\right)^{2}}\\
\beta=\frac{\overline{q^{2}}}{\overline{q^{4}}-\left(\overline{q^{2}}\right)^{2}}\end{array}\right..\label{eq:System3}\end{equation}

In order to use (\ref{eq:System3}), we should require the equivalence
of the first two even moments of (\ref{eq:Gamma}) to those of the
ideal scintillator response, which can be easily calculated with (\ref{MomentN}):

\[
\overline{q^{2}}=\mu^{2}+\mu(1+v_{1});\]

\[
\overline{q^{4}}=\mu(1+6\mu+4\mu^{2}+v_{1}^{2}(3+2\mu)+2v_{1}(3+8\mu+2\mu^{2}))+\left(\overline{q^{2}}\right)^{2}.\]

\end{document}